\documentclass[camera,pageno]{jpaper}

\usepackage{multirow}
\usepackage{mathtools}
\usepackage{xspace}
\usepackage{textcomp}
\usepackage{algorithm}
\usepackage{algorithmic}
\usepackage{subfig}
\usepackage{xfrac}
\usepackage{indentfirst}
\usepackage{subscript}
\usepackage[multiple]{footmisc}

\usepackage{ifthen}
\usepackage{calc}
\usepackage{pifont}
\usepackage{color}
\usepackage{fancyhdr}
\usepackage{tikz}

\newcounter{hours}
\newcounter{minutes}

\newboolean{timeofmake}
\setboolean{timeofmake}{false}

\newcommand{\ignore}[1]{}

\newcommand{\intersub}[0]{{Subarray Access Refresh Parallelization}\xspace}

\newcommand{\is}[0]{{\text{SARP}}\xspace}
\newcommand{\sarp}[0]{{\text{SARP}}\xspace}

\newcommand{\warplongCap}[0]{{\mbox{Write-refresh} Parallelization}\xspace}
\newcommand{\warplong}[0]{{\mbox{write-refresh} parallelization}\xspace}
\newcommand{\Warplong}[0]{{\mbox{Write-refresh} parallelization}\xspace}
\newcommand{\darp}[0]{{DARP}\xspace}
\newcommand{\darplong}[0]{{Dynamic Access Refresh Parallelization}\xspace}

\newcommand{\ooolong}[0]{{\mbox{out-of-order} per-bank refresh}\xspace}
\newcommand{\ooolongCap}[0]{{\mbox{Out-of-order} Per-bank Refresh}\xspace}
\newcommand{\ib}[0]{{\darp}\xspace}

\newcommand{\combo}[0]{{DSARP}\xspace}

\newcommand{\refab}[0]{{\small\emph{$REF_{ab}$}}\xspace}
\newcommand{\refpb}[0]{{\small\emph{$REF_{pb}$}}\xspace}

\newcommand{\caprefab}[0]{\small{\textbf{\textit{REF\textsubscript{ab}}}}\xspace}
\newcommand{\caprefpb}[0]{\small{\textbf{\textit{REF\textsubscript{pb}}}}\xspace}

\newcommand{\figputHW}[3]{
\begin{figure}[h]
\begin{minipage}{\linewidth}
\footnotesize 
\begin{center}
\includegraphics[width=1.0\linewidth]{plots/#1}
\end{center}
\vspace{-0.1in}
\caption{#2 \label{fig:#1}}
\end{minipage}
\end{figure}
}

\newcommand{\figputGHS}[3]{
\begin{figure}[h]
\begin{minipage}{\linewidth}
\begin{center}
\includegraphics[scale=#2]{gnuplots/#1}
\end{center}
\vspace{-0.1in}
\caption{#3 \label{fig:#1}}
\end{minipage}
\end{figure}
}

\newcommand{\figref}[1]{Figure~\ref{fig:#1}}

\begin{document}

\pretitle{\begin{center}\vspace{-.5in}
SAFARI Technical Report No. 2015-001. Feb 20, 2015.\\
This is a summary of the original paper, entitled "Improving DRAM Performance by
Parallelizing Refreshes with Accesses" which appears in HPCA
2014~\cite{chang-hpca2014}.\\
\vspace{0.05in}
\normalfont\Large\bfseries}
\posttitle{\par\end{center}}

\title{Reducing Performance Impact of DRAM Refresh \\by Parallelizing Refreshes with
Accesses}

\preauthor{\begin{center}\large
    \begin{tabular}[t]{ccc}
          Kevin Kai-Wei Chang & Donghyuk Lee & Zeshan Chishti$\dagger$ \\
    \end{tabular}
        \vskip .5em
    \begin{tabular}[t]{cccc}
        Alaa R. Alameldeen$\dagger$ &  Chris Wilkerson$\dagger$ & Yoongu Kim & Onur Mutlu \\
    \end{tabular}
        \vskip .5em
    \begin{tabular}[t]{cc}
        Carnegie Mellon University & $\dagger$Intel Labs \\
}
\postauthor{\end{tabular}\par\end{center}\vspace{-0.2in}}

\date{}
\maketitle

\thispagestyle{empty}
{
    \setstretch{0.93}
    \frenchspacing
    \section{Summary}

\subsection{The Problem}
DRAM requires periodic refresh to prevent data loss from charge leakage. There
exists two main refresh methods employed in the majority of today's DRAM
systems. The first method is to carry out refresh operations at the rank level,
called all-bank refresh (\refab), which is mainly used by commodity DDR
DRAM~\cite{jedec-ddr4}. Because all-bank refresh prevents all banks within an
entire DRAM rank from serving memory requests, it significantly degrades
performance. The second method is to perform refreshes at the bank level, called
per-bank refresh (\refpb), which is currently supported in LPDDR DRAM used in
mobile platforms~\cite{jedec-lpddr3}. In contrast to \refab, \refpb enables a
bank to be accessed while another bank is being refreshed, alleviating part of
the negative performance impact of refresh.

Unfortunately, there are two shortcomings of per-bank refresh. First, refreshes
to different banks are scheduled in a strict round-robin order as specified by
the LPDDR standard~\cite{jedec-lpddr3}. Using this static policy may force a
busy bank to be refreshed, delaying the memory requests queued in that bank,
while other idle banks are available to be refreshed.
Second, refreshing banks cannot concurrently serve
memory requests. Furthermore, the negative performance impact of DRAM refresh
becomes exacerbated as DRAM density increases in the future.
\figref{perf_loss_all} shows the average performance degradation of
all-bank/per-bank refresh compared to ideal baseline without any
refreshes.\footnote{The detailed methodology is described in our HPCA
paper~\cite{chang-hpca2014}.} Although \refpb performs slightly better than
\refab, the performance loss is still significant, especially as the density
grows (16.6\% loss at 32Gb). Therefore, \textbf{the goal} of our
paper~\cite{chang-hpca2014} is to provide practical mechanisms to overcome these
two shortcomings to mitigate the performance overhead of DRAM refresh.

\figputGHS{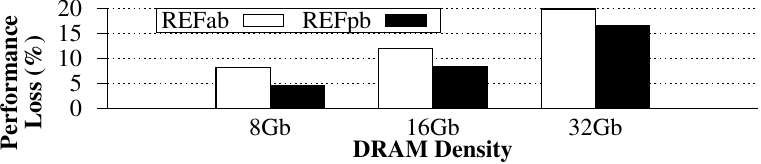}{1.0}{Performance loss due to \caprefab and \caprefpb.}

\subsection{Proposed Solutions}

We propose two mechanisms, \emph{\darplong (\darp)} and \emph{\intersub (\is)},
that hide refresh latency by parallelizing refreshes with memory accesses across
\emph{banks} and \emph{subarrays}, respectively. \darp is a new refresh
scheduling policy that consists of two components. The first component is
\emph{\ooolong} that enables the memory controller to specify a particular
(idle) bank to be refreshed as opposed to the standard per-bank refresh policy
that refreshes banks in a strict round-robin order. With out-of-order refresh
scheduling, \darp can avoid refreshing (non-idle) banks with pending memory
requests, thereby avoiding the refresh latency for those requests. The second
component is \emph{\warplong} that proactively issues \refpb to a bank
while DRAM is draining write batches to other banks, thereby overlapping refresh
latency with write latency. The second mechanism, \is, allows a bank to serve
memory accesses in idle subarrays while other subarrays within the same bank are
being refreshed. \is exploits the fact that refreshing a row is contained within
a subarray, without affecting the I/O bus used for transferring data.

\subsubsection{\darp: \ooolongCap} The limitation of the current \refpb mechanism is
that it disallows a memory controller from specifying which bank to refresh.
Instead, a DRAM chip has internal logic that strictly refreshes banks in a
\emph{sequential round-robin order}. Because DRAM lacks visibility into a memory
controller's state (e.g., request queues' occupancy), simply using an in-order
\refpb policy can unnecessarily refresh a bank that has multiple pending
requests to be served when other banks may be free to serve a refresh command.
To address this problem, we propose the first component of \darp,
\emph{\ooolong}. The idea is to remove the bank selection logic from DRAM and
make it the memory controller's responsibility to determine which bank to
refresh. As a result, the memory controller can refresh an idle bank to enhance
parallelization of refreshes and accesses, avoiding refreshing a bank that has
pending requests as much as possible.

Due to \refpb reordering, the memory controller needs to guarantee that
deviating from the original in-order schedule still preserves data integrity. To
achieve this, we take advantage of the fact that the contemporary DDR JEDEC
standard~\cite{jedec-ddr4} provides some refresh scheduling
flexibility. The standard allows up to \emph{eight} all-bank refresh commands to
be issued late (postponed) or early (pulled-in). This implies that each bank can
tolerate up to eight \refpb to be postponed or pulled-in. Therefore, the memory
controller ensures that reordering \refpb preserves data integrity by limiting
the number of postponed or pulled-in commands. Our paper~\cite{chang-hpca2014}
describes the algorithm of \ooolong in detail.

\subsubsection{\darp: \warplongCap} The key idea of the second component of
\darp is to actively avoid refresh interference on read requests and instead
enable more parallelization of refreshes with \emph{write requests}. We make two
observations that lead to our idea. First, {\em write batching} in DRAM creates
an opportunity to overlap a refresh operation with a sequence of writes, without
interfering with reads. A modern memory controller typically buffers DRAM writes
and drains them to DRAM in a batch to amortize the \emph{bus turnaround
latency}, also called \emph{tWTR} or
\emph{tRTW}~\cite{jedec-ddr4,kim-isca2012,lee-tech2010}, which is the additional
latency incurred from switching between serving writes to reads. Typical systems
start draining writes when the write buffer occupancy exceeds a certain
threshold until the buffer reaches a low watermark. This draining time period is
called the \emph{writeback mode}, during which no rank within the draining
channel can serve read
requests~\cite{chatterjee-hpca2012,lee-tech2010,stuecheli-isca2010}. Second,
DRAM writes are not latency-critical because processors do not stall to wait for
them: DRAM writes are due to dirty cache line evictions from the last-level
cache~\cite{lee-tech2010,stuecheli-isca2010}.

Given that writes are not latency-critical and are drained in a batch for some
time interval, they are more flexible to be scheduled with minimal performance
impact. We propose the second component of \darp, \emph{\warplong}, that
attempts to maximize parallelization of refreshes and writes. \Warplong selects
the bank with the minimum number of pending demand requests (both read and
write) and preempts the bank's writes with a per-bank refresh. As a result, the
bank's refresh operation is hidden by the writes in other banks. We refer the
reader to Section 4 of our paper~\cite{chang-hpca2014} for more details on the
algorithm and implementation of \darp.

\subsubsection{\intersub (SARP)} To tackle the problem of refreshes and accesses colliding
within the same bank, we propose \emph{\sarp (\intersub)} that exploits the
existence of subarrays within a bank. The key observation leading to our second
mechanism is that a refresh operation is constrained to only a few \emph{subarrays} within a bank
whereas the other \emph{subarrays} and the \emph{I/O bus} remain idle during the
process of refreshing. The reasons for this are two-fold.  First, refreshing a
row requires only its subarray's sense amplifiers that restore the charge in the
row without transferring any data through the I/O bus.  Second, each subarray
has its own set of \emph{sense amplifiers} that are not shared with other
subarrays.

Based on this observation, SARP's key idea is to allow memory accesses to an
\emph{idle} subarray while other subarrays are refreshing.
\figref{subarray-service-timeline} shows the service timeline and the
performance benefit of our mechanism. As shown, \is reduces the read latency by
performing the read operation to Subarray 1 in parallel with the refresh in
Subarray 0. Compared to \ib, \is provides the following advantages: 1) \is is
applicable to both all-bank and per-bank refresh and 2) \is enables memory accesses
to a refreshing bank, which cannot be achieved with \ib.


\figputHW{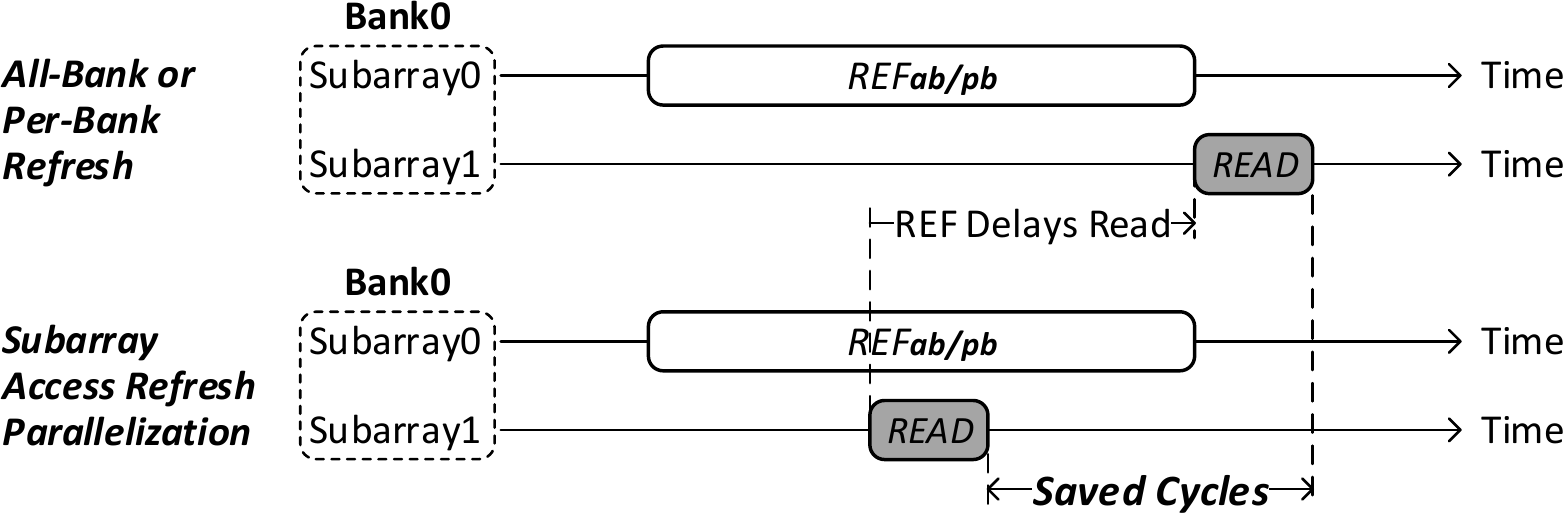}{Service timeline of a refresh and a read request to two
different subarrays within the same bank.}

\is requires modest modifications that reduce the sharing of the peripheral
circuits for refreshes and accesses in each bank without changing the cell
arrays. These modifications result in 0.71\% DRAM die area overhead. Section 4.3
of our paper~\cite{chang-hpca2014} describes these changes in detail.


\subsection{Summary of Results}

%
Here, we briefly summarize our results on an 8-core system. Section 6 of our
paper provides the detailed evaluations. \figref{toppick-results} shows the
average system performance and energy of our final mechanism, \combo, the
combination of \darp and \sarp, compared to two baseline refresh schemes and an
ideal scheme without any refreshes. The percentage numbers on top of the bars
are the performance improvement of \combo over \refab. We draw two observations.
First, \combo consistently improves system performance and energy efficiency
over prior refresh schemes, capturing most of the benefit of the ideal baseline.
Second, as DRAM density (refresh latency) increases, the performance benefit of
\combo gets larger.

\figputGHS{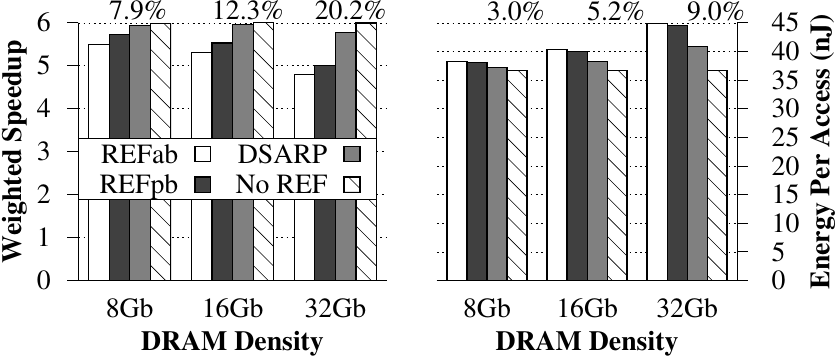}{1.0}{Average system performance and energy consumption.}

    \section{Significance}

\subsection{Novelty} To our knowledge, this is the first work to comprehensively
study the effect of per-bank refresh and propose 1) a refresh scheduling policy
built on top of per-bank refresh and 2) a mechanism that achieves
parallelization of refresh and memory accesses {\em within} a refreshing bank.
As a result, our mechanisms significantly improve system performance by
effectively parallelizing refreshes with accesses.

Prior works have investigated refresh scheduling policies on all-bank refresh or
DDR4 fine granularity refresh to hide refresh operations behind rank idle
time~\cite{mukundan-isca2013,stuecheli-micro2010}.\footnote{A recent
work~\cite{tavva-taco2014} that was published after our work proposes to operate
DDR4 refresh at sub-rank level, which performs worse than per-bank refresh.}
However, they provide minimal performance benefits because an all-bank refresh
operates at the \emph{rank level}, occupying all the banks, thus making it
difficult to find a long enough idle period to hide the refresh latency. On the
other hand, some works have proposed to exploit retention time variation among
DRAM cells~\cite{liu-isca2012,venkatesan-hpca2006}. Although this approach has
the potential to reduce the number of refreshes, determining the retention time
of DRAM cells accurately is still an unsolved research problem due to the
Variable Retention Time and Data Pattern Dependence
phenomena~\cite{khan-sigmetrics2014,liu-isca2013}. In comparison, our proposed
techniques do not rely on retention time profiling and are guaranteed to
preserve data integrity.

Other works have proposed to skip refreshes in different scenarios. Ghosh and
Lee~\cite{ghosh-micro2007} propose to skip refreshes to rows that had been
recently accessed.  Liu et al.~\cite{liu-asplos2011} propose Flikker to lower
refresh rate of non-critical data regions. Isen and John~\cite{isen-micro2009}
propose ESKIMO to avoid refreshing invalid or unused data based on program
semantics. DSARP is complementary to these techniques.




\subsection{Potential Long Term Impact}

In this section, we describe three trends in the current and future DRAM
subsystem that will likely make our proposed solutions more important in the
future.

\subsubsection{Worsening retention time} As DRAM cells' feature size continues
to scale, the cells' retention time will likely become shorter,
exacerbating the refresh penalty~\cite{mutlu-imw2013}. When the surface area of cells gets smaller
with further scaling, the depth/height of the cell needs to increase to maintain
the same amount of capacitance that can be stored in a cell. In other words, the
\emph{aspect ratio} (the ratio of a cell's depth to its diameter) needs to be
increased to maintain the capacitance. However, many works have shown that
fabricating high aspect-ratio cells is becoming more difficult due to processing
technology~\cite{hong-iedm2010, kang-2014}. Therefore, cells' capacitance
(retention time) may potentially decrease with further scaling, increasing the
refresh frequency. Using DSARP is a cost-effective way to alleviate the
increasing negative impact of refresh as our results show~\cite{chang-hpca2014}.


\subsubsection{New DRAM standards with flexible per-bank refresh} According to
the recently released DRAM standards, the industry is already in the process of
implementing a similar concept of enabling the memory controller to determine
which bank to refresh. In particular, the two standards are: 1)
HBM~\cite{jedec-hbm} (October 2013, after the submission of our work) and 2)
LPDDR4~\cite{jedec-lpddr4} (August 2014). Both standards have incorporated a new
refresh mode that allows per-bank refresh commands to be issued in any order by
the memory controllers. Neither standard specifies a preferred order which the
memory controller needs to follow for issuing refresh commands.

Our work has done extensive evaluations to show that our proposed per-bank
refresh scheduling policy, \emph{DARP}, outperforms a naive round-robin
policy by opportunistically refreshing idle banks. As a result, our policy can
be potentially adopted in the future processors that use HBM or LPDDR4 DRAM.

\subsubsection{Increasing number of subarrays} As DRAM density keeps increasing,
more rows of cells are added within each DRAM bank. To avoid the disadvantage of
increasing sensing latency due to longer bitlines in subarrays, more subarrays
will likely be added within a single bank instead of increasing the size of each
subarray. Our proposed refreshing scheme at the subarray level, \emph{SARP},
becomes more effective at mitigating refresh as the number of subarrays
increases because the probability of a refresh and a demand request colliding
at the subarray level decreases with more subarrays.

\subsection{New Research Directions}

This work will likely create new research opportunities for studying refresh scheduling
policies at different dimensions (i.e., bank and subarray level) to mitigate worsening
refresh overhead.
Among many potential opportunities, one potential way to further reduce the
refresh latency (i.e., {\small\emph{$tRFC_{ab/pb}$}}) is to trade off higher
refresh rate (i.e., {\small\emph{$tREFI$}}), which is currently supported as
\emph{fine granularity refresh} in DDR4 DRAM for all-bank refresh. In this work,
we assume a fixed refresh rate for per-bank refresh as it is specified in the
standard. Therefore, a new research question that our work raises is \emph{how
can one combine per-bank refresh with fine granularity refresh and design a
new scheduling policy for that}? We think that \darp can inspire new scheduling
policies to improve the performance of existing DRAM designs.

\ignore{An important research question that existing works do not discuss is
\emph{how can one design a QoS-aware refresh scheduling policy to provide QoS
and high performance?} Because refreshing a row of data essentially reads that
row into its row buffer, it evicts the previous row stored in the same row
buffer, increasing the memory latency of subsequent accesses. As a result,
designing refresh policies to avoid causing row misses to QoS-critical
applications is an important research direction.}



{
\vspace{-0.4mm}
\small
\bstctlcite{bstctl:etal, bstctl:nodash, bstctl:simpurl}
\bibliographystyle{IEEEtranS}
\bibliography{paper}
}
}

\end{document}